\documentclass[a4paper,11pt]{article}
\usepackage{pos}
\usepackage[symbol]{footmisc}

\title{Galactic molecular clouds as sources of secondary positrons}

\author*[a]{Agnibha De Sarkar}
\emailAdd{agnibha@rri.res.in}
\author[a]{Sayan Biswas}
\emailAdd{sayan@rri.res.in}
\author[a]{Nayantara Gupta}
\emailAdd{nayan@rri.res.in}

\affiliation[a]{Raman Research Institute,\\
C. V. Raman Avenue, 5th Cross Road, Sadashivanagar, Near Mekhri Circle, Bengaluru, Karnataka 560080}

\abstract{Secondary positrons produced inside Galactic Molecular Clouds (GMCs) can significantly contribute to the observed positron spectrum on Earth. Multi-wavelength data of GMCs are particularly useful in building this model. A very recent survey implemented the optical\,/\,IR dust extinction measurements to trace 567 GMCs within 4 kpc of Earth, residing in the Galactic plane. We use the updated catalog of GMCs reported in recent papers, distributed in the Galactic plane, to find the secondary positrons produced in them in interactions of cosmic rays with molecular hydrogen. Moreover, by analyzing the \textit{Fermi}-LAT data, new GMCs have been discovered near the Galactic plane. We also include some of these GMCs closest to the Earth, where cosmic ray interactions produce secondaries. It has been speculated earlier that cosmic rays may be reaccelerated in some GMCs. We select 7 GMCs out of 567 GMCs recently reported, within 4 kpc of Earth, where reacceleration due to magnetized turbulence is assumed. We include a hardened component of secondary positrons produced from the interaction of reaccelerated CRs in those 7 GMCs. We use publicly available code \texttt{DRAGON} for our simulation setup to study CR propagation in the Galaxy and show that the observed positron spectrum can be well explained in the energy range of 1 to 1000 GeV by our self-consistent model.}

\FullConference{37$^{\rm{th}}$ International Cosmic Ray Conference (ICRC 2021)\\
		July 12th -- 23rd, 2021\\
		Online -- Berlin, Germany}

\begin{document}
\maketitle

\section{Introduction}
Recent progress in space-based and ground-based experiments has helped measure fluxes of cosmic ray (CR) nuclei, electron (e$^-$), positron (e$^+$), antiproton ($\bar{p}$) etc with unprecedented precision. Especially, AMS-02 aboard International Space Station, has provided accurate measurements of e$^-$, e$^+$, e$^+$/(e$^-$+e$^+$), in the energy range 0.1 GeV to 1 TeV. Such measurements of different observables have helped elucidate interesting properties related to large scale CR propagation in the Milky Way Galaxy. One such phenomenon is the so called "Positron excess" problem, which is recently confimed by AMS-02 experiment \cite{aguilar19b}. This was also confirmed by PAMELA experiment \cite{adriani13dataa}. Positron excess is a phenomena in which the observed positron flux  rises with energy and shows a peak near a few hundred GeV, and then a subsequent drop-off is observed. This phenomena pose a serious challenge to the scientists, as this can not be explained by the standard paradigm of  interactions of CRs with interstellar hydrogen gas. Different innovative scenarios such as Pulsars, dark matter, microquasar jets were implemented as the origin of positron excess. However, it has not been conclusively confirmed what actually creates this excess.

Here, we have considered an alternative approach to explain the positron excess. Giant Molecular Clouds (GMCs) are dense reservoirs of cold protons, distributed throughout the Galactic plane. Such concentrated clumps of protons can be ideal laboratory for different particle interactions. Considering Galactic SNRs as primary sources of CRs, we have constructed a self-consistent model using publicly available code \texttt{DRAGON} \footnote{The 3D version of the DRAGON code is available at
https://github.com/cosmicrays/DRAGON for download.}, in which we have considered interactions of primary CRs with interstellar hydrogen gas (CASE 1), interactions inside GMCs residing on the Galactic plane \cite{rice16,chen20,aharonian16} (CASE 2) and contribution from nearby, sub-kpc GMCs \cite{chen20,aharonian16} (CASE 3). In earlier studies, extensive theoretical work has shown that primary CRs can get reaccelerated inside GMCs due to magnetized turbulence \cite{dogiel87}. As a result of which, CR spectra and in turn, spectra of secondary positrons produced from the interaction, will get hardened. Since this spectral hardening inside GMCs is yet to be observed conclusively, we have imposed three conditions i.e. 1) detection incapability of Fermi-LAT, 2) radius of the GMCs $\ge$ 10 pc and 3)  distance from the Earth $\le$ 1 kpc, to select 7 GMCs from \cite{chen20}. We have assumed that reacceleration of primary CRs due to turbulence occurs in these selected nearby 7 GMCs and added their contribution to the total lepton flux (CASE 3). Finally we show that  the total flux from CASE 1, CASE 2 and CASE 3 can explain the positron excess observed by AMS-02 and PAMELA data  in 1 GeV to 1 TeV energy range. Our self-consistent model also satisfies various CR observables such as proton, antiproton spectra and B/C and $^{10}$Be\,/\,$^9$Be ratios quite well. We refer to the principal paper for the extended version of the results \cite{agnibha21}.

\section{Propagation model setup}

In this work, we study the CR propagation by solving the CR transport equation numerically, using \texttt{DRAGON} code. \texttt{DRAGON} extensively captures various physical processes such as propagation and scattering of CRs in regular and turbulent magnetic fields, CRs interacting with ISM and GMCs, energy losses due to radioactive decay of the nuclei, ionization loss, Coulomb loss, Bremsstrahlung loss, synchrotron and IC loss, re-acceleration and convection in the Galactic medium, to obtain the solution of the transport equation for the CR propagation in the Galaxy. We have solved transport equation in 3D geometry, assuming the Galaxy to be cylindrical in shape. The observer is set at Sun's position x = 8.3 kpc, y = 0 kpc, z = 0 kpc. We consider outermost radial boundary of the Galaxy R$_{max}$ = 25 kpc, halo height z$_t$ = 8 kpc and vertical boundary L = 3z$_t$.

\begin{figure}
\includegraphics[width=0.5\textwidth,origin=c,angle=0]{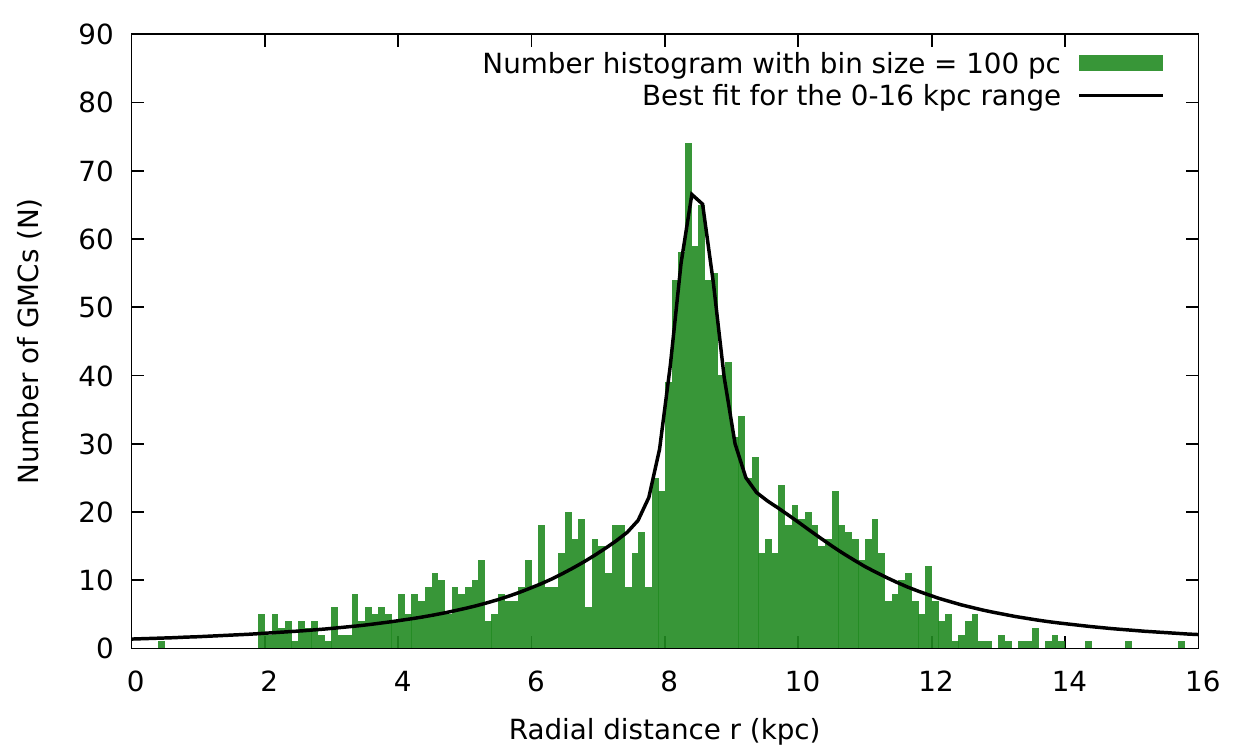}
\includegraphics[width=0.5\textwidth,origin=c,angle=0]{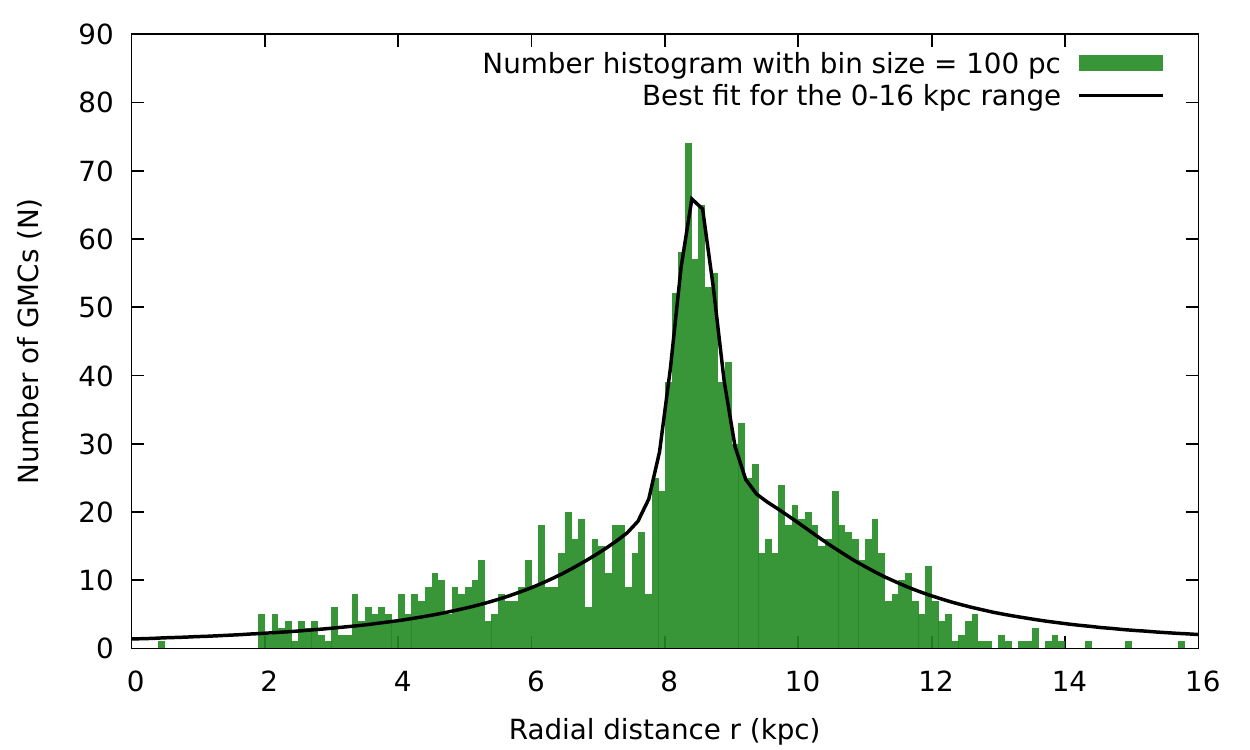}
\caption{\label{fig1} \textit{Left :} Histogram 1 : GMCs considered from \cite{rice16,chen20,aharonian16} (CASE 2 + CASE 3) and \textit{Right :} Histogram 2 : GMCs considered from \cite{rice16,chen20,aharonian16} (CASE 2), apart from 10 nearby GMCs considered for CASE 3. The black line signifies the functional fit of the histogram in the entire spatial range.}
\end{figure}

We consider Galactic SNRs as major sources of primary CRs with an universal injection spectrum, corresponding to work done by \cite{ferriere01}. The Galactic magnetic field model was taken from \cite{pshirkov11}. The regular magnetic field was considered to be in the range 2-11 $\mu$G \cite{pshirkov11} and turbulent magnetic field component was calculated using the expression given in \cite{bernardo13}. Isotropic diffusion was considered with exponential vertical dependence, in the form of D($\rho$,\,z) = $\beta^{\eta}\,$D$_0\,(\frac{\rho}{\rho_0})^{\delta}$\,exp($\frac{z}{z_t}$), where $\rho$ is the rigidity, $\delta$ is the index, $\beta$ is a dimensionless particle velocity and D$_0$ = D($\rho_0$) is the normalization at reference rigidity $\rho_0$ = 4 GV. The atomic, ionized hydrogen gas distribution and interstellar radiation field (ISRF) were fixed on the basis of astronomical data. To construct the molecular hydrogen distribution, we consider latest catalog of 1064 molecular clouds from \cite{rice16}, which was built using dendogram-based decomposition of a previous most uniform, large-scale all-Galaxy CO survey \cite{dame01}. Next, we consider 567 GMCs observed within 4 kpc from Earth traced by optical/near-infrared (NIR) dust extinction measurements \cite{chen20}. Additionally, we take into account nearby GMCs reported by \cite{aharonian16}, in which the authors have analyzed \textit{Fermi}-LAT gamma ray data of these GMCs. We add all of these GMCs and fit the radial number distribution N(r) with a smooth pseudo-voigt profile i.e. a linear combination of Gaussian and Lorentzian profiles. Assuming average density of the distribution <n$_{H_2}$> $\sim$ 100 cm$^{-3}$, we calculate molecular hydrogen distribution by the equation n$_{H_2}$ (r) = <n$_{H_2}$> $\times$ $\left(\frac{N(r)}{N_{total}}\right)$. For fitting proton, antiproton spectra and B/C, $^{10}$Be/$^{9}$Be ratios, we have considered all of the GMCs in the number distribution i.e. (CASE 2 + CASE 3) (Histogram 1). However, since high energy leptons lose energy radiatively very fast with distance, nearby lepton sources should be treated separately. That is why, for fitting lepton spectra and lepton fraction, we have used all of the GMCs in the number distribution i.e. CASE 2 (Histogram 2), apart from 3 GMCs from \cite{aharonian16} and 7 GMCs from \cite{chen20}, where reacceleration is assumed. Contribution from these 10 GMCs were calculated separately in CASE 3. The histograms are shown in Figure \ref{fig1}.

In this work, we have used broken power law as injection spectra of protons and heavy nuclei (up to Z = 14). For primary electrons, we have used similar broken power law as injection spectra. Modulation of CR spectra due to solar influence was modelled using force-field approximation according to \cite{usoskin05}. We have primarily used AMS-02 and PAMELA data to fit our model in this work.

\section{Nearby GMCs contribution}

After combining the contribution from CASE 1 and CASE 2, expectedly total calculated positron flux does not satisfy the observed flux by AMS-02 and PAMELA. Hence, we take contribution from nearby GMCs (d $\le$ 1 kpc), which is the CASE 3. Primary CRs injected from SNRs propagate through these clumps of cold protons, subsequently creating gamma ray and leptons through hadronic \textit{p-p} interaction. We also consider reacceleration of CRs due to magnetized turbulence in some of the nearby GMCs, which was first proposed by \cite{dogiel87}.

Taurus, Lupus and Orion A are three closest GMCs from Earth, for which gamma ray data has been analyzed by \cite{aharonian16}. The parent proton density spectrum that is responsible for gamma ray emission from these clouds, has been found to be in a power law form J$_p$(E$_p$) = $\rho_{0, CR}\left(\frac{E_p}{E_0}\right)^{-\alpha}$, where $\rho_{0, CR}$ is the normalization constant, E$_0$ is the reference energy and $\alpha$ is the spectral index. Using the parameter values (and also considering the uncertainty ranges) of the parent proton spectrum, which has been worked out from the gamma ray observation, we calculate the secondary leptons produced using the formalism given in \cite{kelner06}.

Although it has not been observationally confirmed yet, \cite{dogiel87} has argued that particle energies may increase or reacceleration may happen due to fluctuating electro-magnetic (EM) field inside the GMCs. Since energy of the turbulence due to fluctuating EM field is comparable to gravitational energy of the cloud, in a cloud-collapse scenario, the turbulence inside GMCs can slow down the cloud from collapsing due its own gravitation. So part of the gravitational energy then transforms into turbulence energy. \cite{dogiel87} argued of a theoretical mechanism in which this turbulence energy can then transform into particle energies, which would lead to an effective acceleration of particles inside the GMCs. We have selected 7 nearby GMCs, in which we have assumed reacceleration occurs. However, since this phenomena has not been observed yet, we had to select these GMCs very carefully. We have imposed three strict conditions, based on which we have selected these GMCs, inside which reacceleration is assumed. These conditions are,

\begin{itemize}
\item It may happen that reacceleration is occuring inside GMCs, but they are not detected by \textit{Fermi}-LAT. The value of B parameter, defined as B = M$_5$\,/\,d$^2_{kpc}$, must be less than 0.2 for all the selected GMCs, so that these GMCs are outside of \textit{Fermi}-LAT detection threshold. 
\item The size of the GMCs must be greater than the maximum scale length of plasma turbulence inside GMCs, which in turn reaccelerate the particles to increased energies. The scale length of plasma turbulence required to reaccelerate particles is $\approx$ 10 pc, so we have selected GMCs with radius larger than that value.
\item Since leptons lose energy radiatively very fast, these GMCs must be near the Earth. Hence we have selected GMCs within 1 kpc from Earth, so that their contribution to the observed lepton flux is significant.
\end{itemize}

On the basis of aforementioned conditions, we have selected 7 GMCs from \cite{chen20}, in which we have assumed reacceleration is occuring. \cite{dogiel87} has showed that hardened spectral index of injected positron from these GMCs is $\approx$ -1.7. We use the formalism developed by \cite{atoyan95}, to calculate total lepton flux observed from Taurus, Lupus and Orion A and also these 7 selected GMCs in a continuous injection scenario, and try to fit the observed positron excess. We also fit other CR observables self-consistently with the model developed by us. 

\section{Results}

\begin{figure}
\includegraphics[width=0.5\textwidth,origin=c,angle=0]{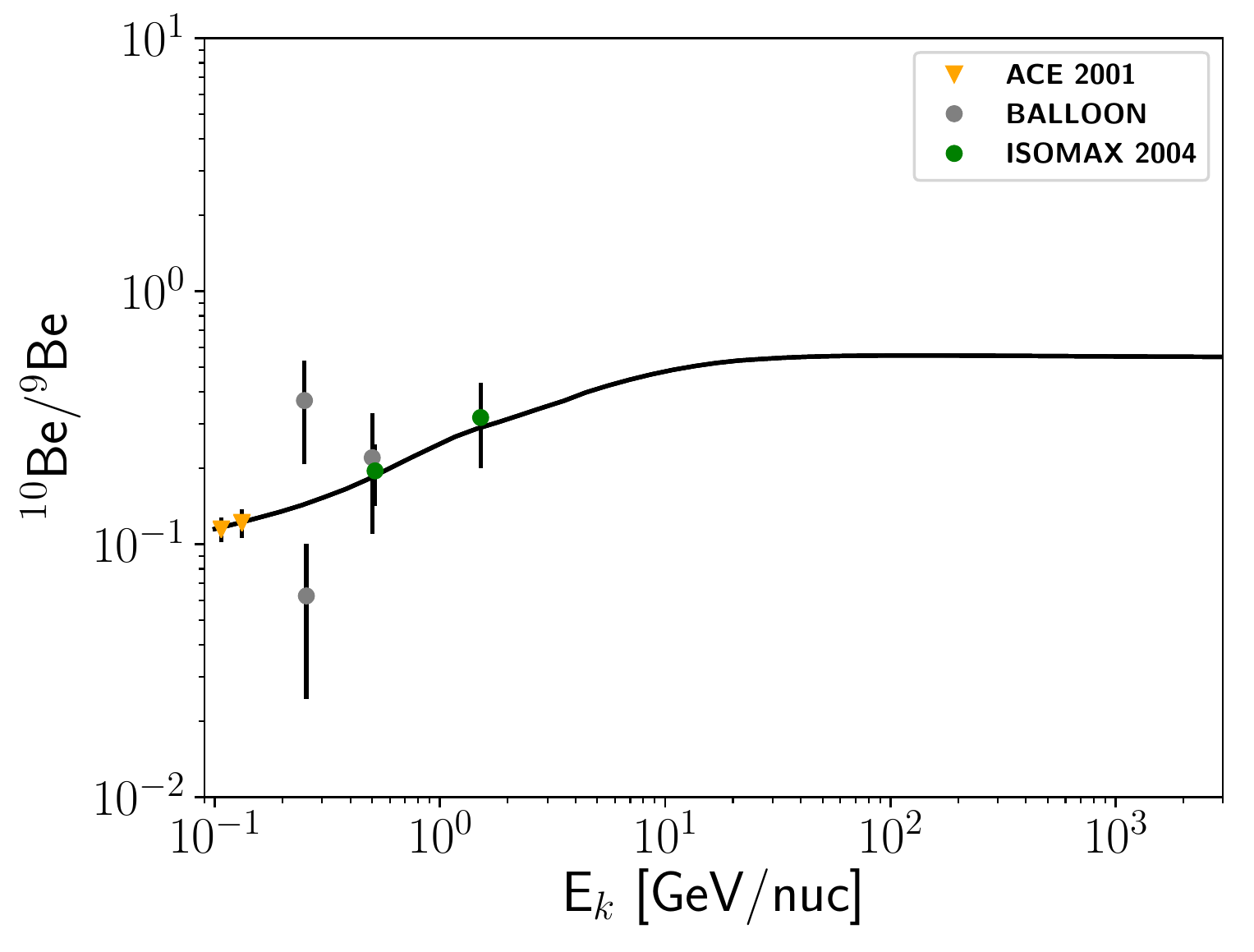}
\includegraphics[width=0.5\textwidth,origin=c,angle=0]{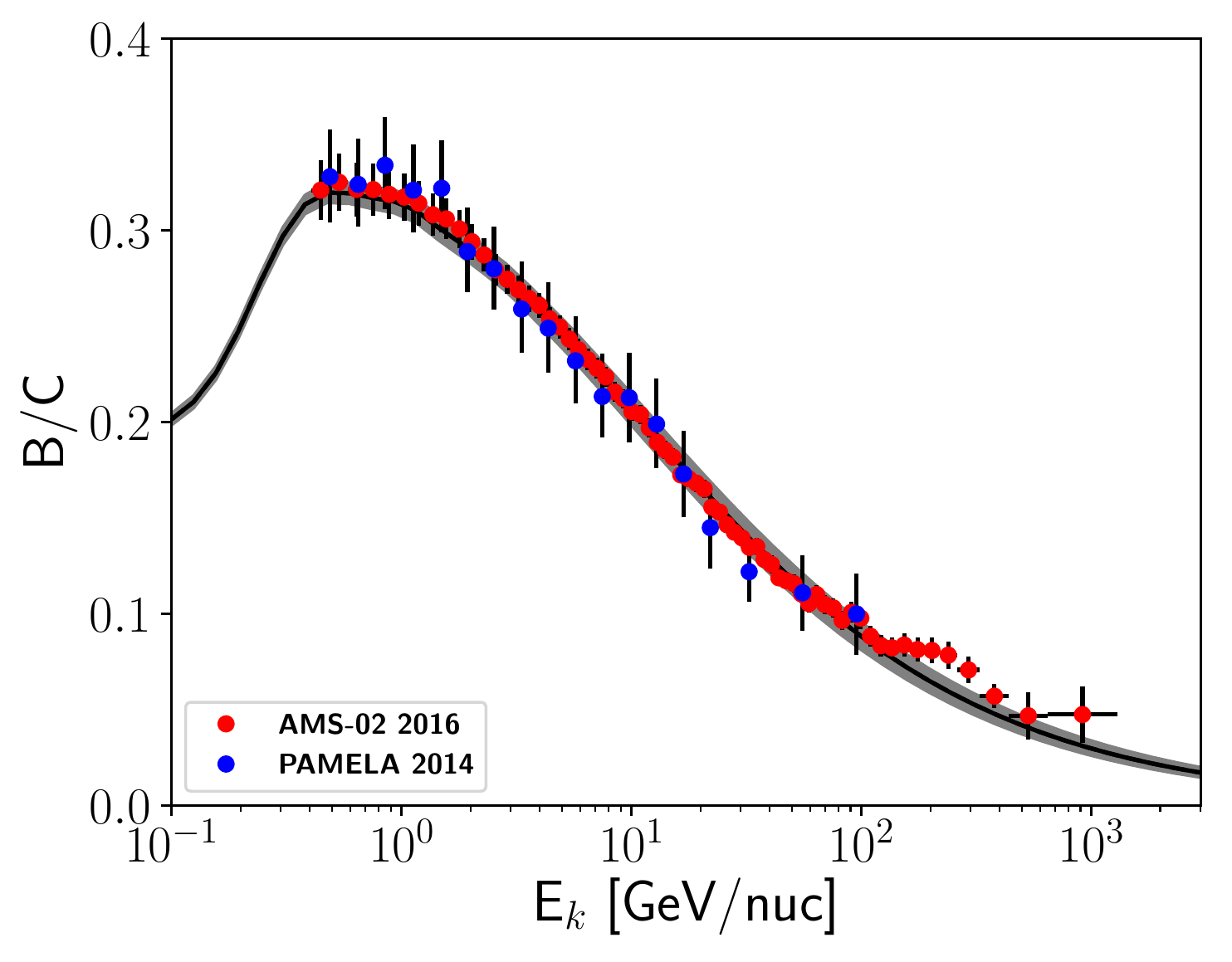}
\caption{\label{fig2} \textit{Left :} $^{10}$Be\,/\,$^9$Be fit from our model against ACE \cite{ACE01}, Ballon and ISOMAX \cite{ISOMAX04} data and \textit{Right :} B\,/\,C fit from our model against AMS-02 \cite{aguilar16a} and PAMELA \cite{adriani14} data. The gray shaded region signifies uncertainties due to variation of propagation parameters.}
\end{figure}

Since $^{10}$Be\,/\,$^9$Be $\propto$ $\sqrt{D(E_k)}$\,/\,z$_t$ and B\,/\,C $\propto$ z$_t$\,/\,D(E$_k$), where E$_k$ is the kinetic energy per nucleon, we can probe the halo height and diffusion coefficient of the Galaxy, by fitting these ratios. In general, $^{10}$Be\,/\,$^9$Be is the possible probe for the halo height and 
B\,/\,C ratio gives us information about the diffusion coefficient in the Galaxy. By fitting $^{10}$Be\,/\,$^9$Be ratio we find that our assumption for halo height to be 8 kpc, is correct. By fitting B\,/\,C ratio, we find the best fit values of the diffusion coefficient to be D$_0$ = 2.4 $\times$ 10$^{29}$ cm$^2$/s, $\delta$ = 0.53, $\eta$ = -0.4, which closely match the standard values for these parameters. The fitted plots are shown in Figure \ref{fig2}.

\begin{figure}
\includegraphics[width=0.5\textwidth,origin=c,angle=0]{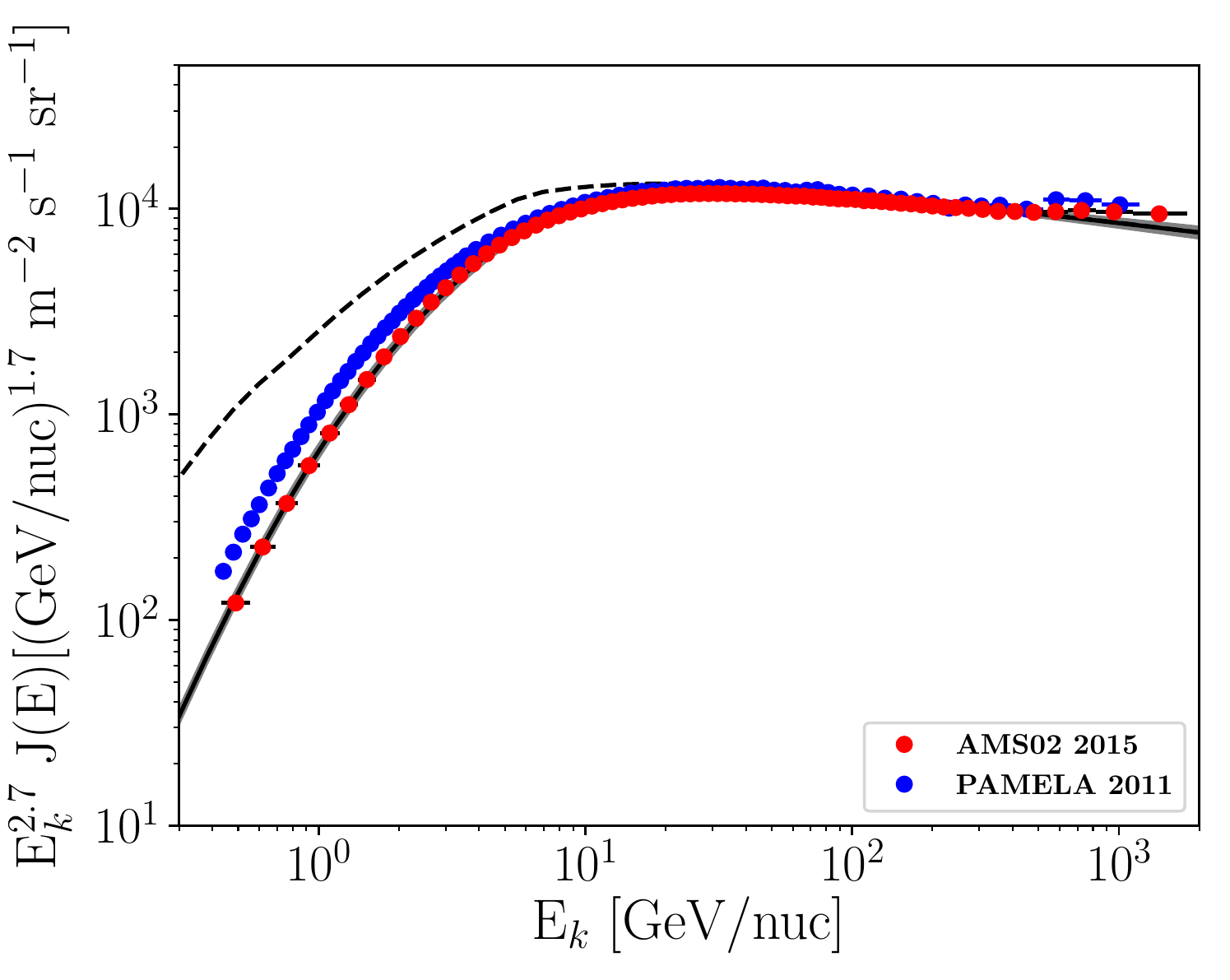}
\includegraphics[width=0.5\textwidth,origin=c,angle=0]{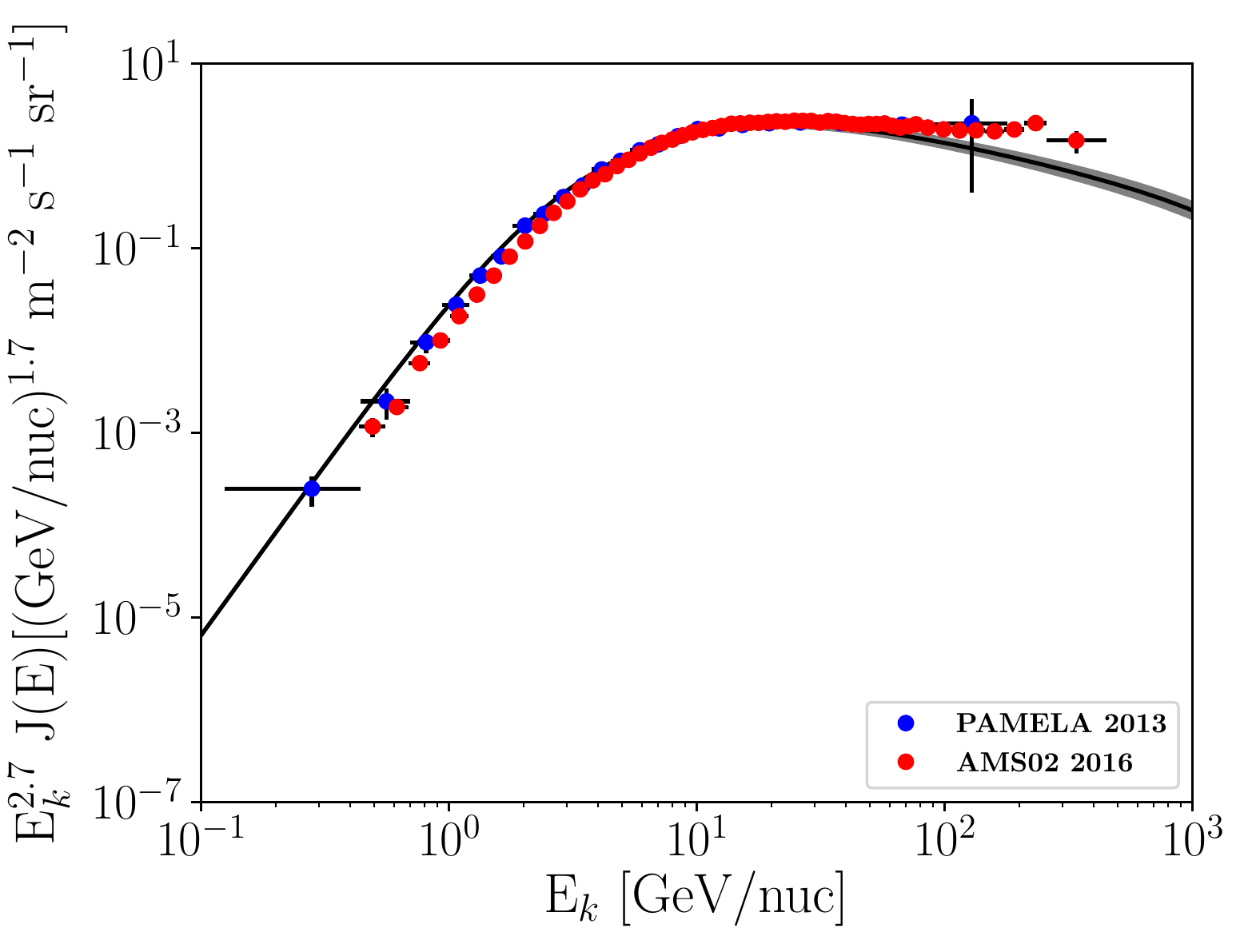}
\caption{\label{fig3} \textit{Left :} Proton spectra fit from our model against AMS-02 \cite{aguilar15a} and PAMELA \cite{adriani11dataa} data and \textit{Right :} Antiproton spectra fit from our model against AMS-02 \cite{aguilar15a} and PAMELA \cite{adriani11dataa} data. The gray shaded region signifies uncertainties due to variation of propagation parameters.}
\end{figure}

Considering the aforementioned halo height and diffusion coefficient parameters, we adjust the injection spectral indices of proton and heavy nuclei to fit the AMS-02 and PAMELA proton and antiproton data. The best-fit injection spectral indices are given as 1.95/2.33, having a spectral break at 7 GV. The plots for the fitted proton and antiproton spectra is shown in Figure \ref{fig3}.

After fitting the hadronic species, we finally try to fit the electron and positron spectra. First, we have adjusted the primary electron spectra to try and fit the electron and positron spectra, considering only CASE 1 + CASE 2. The injection spectral indices of primary electron is given as 2.0/2.7/2.4, with two spectral breaks at 8 and 65 GV. An energy cutoff was also implemented in the electron injection spectra at 10 TeV. As expected, we could not fit electron and positron data satisfactorily. Next we add total lepton contribution from nearby GMCs Taurus, Lupus and Orion A, as well as 7 selected GMCs where reacceleration due magnetized turbulence was assumed i.e. CASE 3 \footnote{See the main paper \cite{agnibha21} for parameter values.}. We show that after taking all of the contribution considered in this work, i.e. CASE 1 + CASE 2 + CASE 3, the positron excess, as well as electron spectum and positron fraction are explained very well. We also calculate anisotropies due to these nearby clouds and show that their anisotropies do not violate that measured by \textit{Fermi-}LAT. We also compare calculated anisotropies of the nearby GMCs with that of nearby pulsars that are usually considered to explain the observed positron excess. We show that we can discern between the contributions from GMC and pulsar scenarios to the positron excess, in terms of observed anisotropy.   

\begin{figure}
\includegraphics[width=0.5\textwidth,origin=c,angle=0]{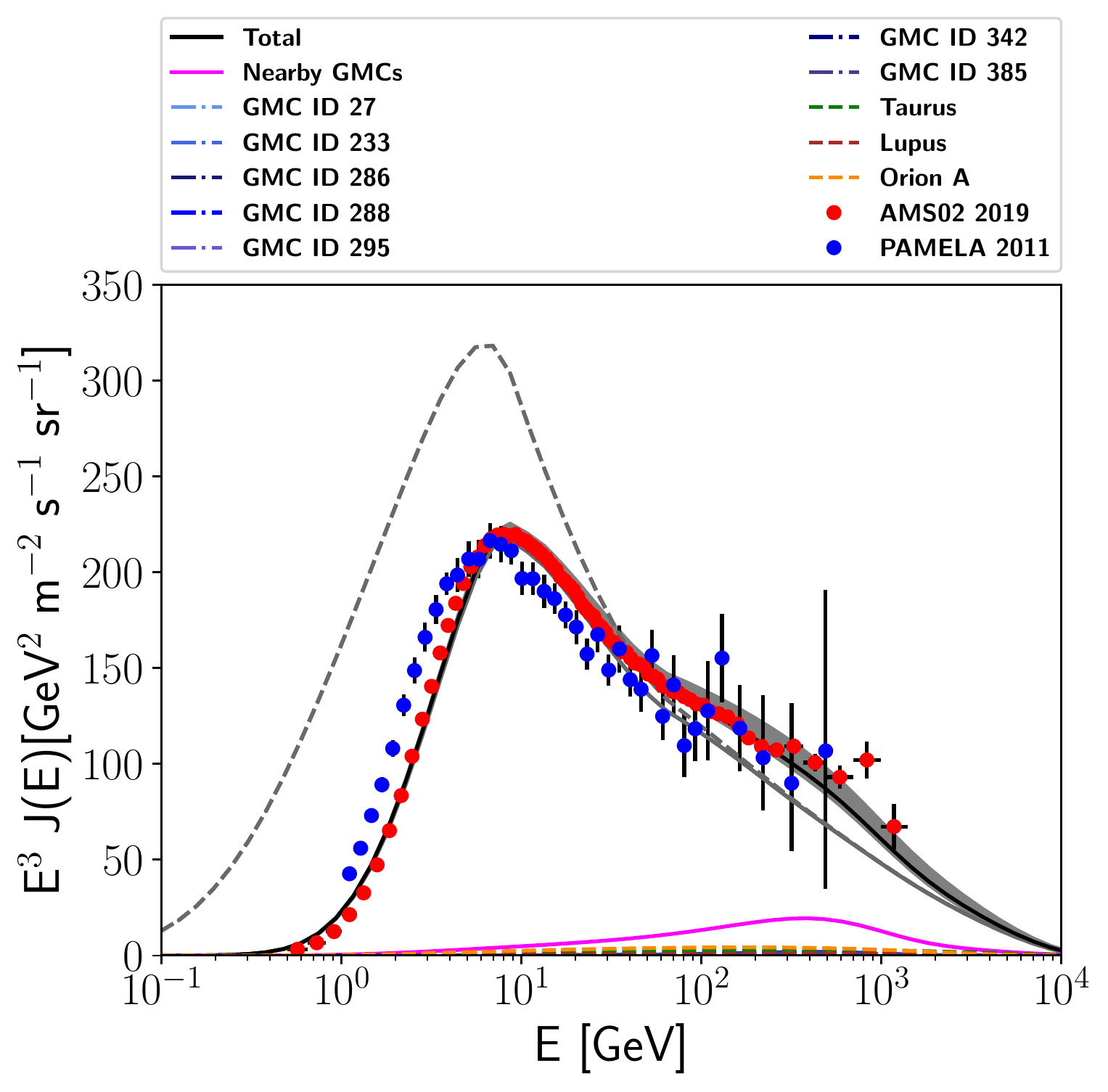}
\includegraphics[width=0.5\textwidth,origin=c,angle=0]{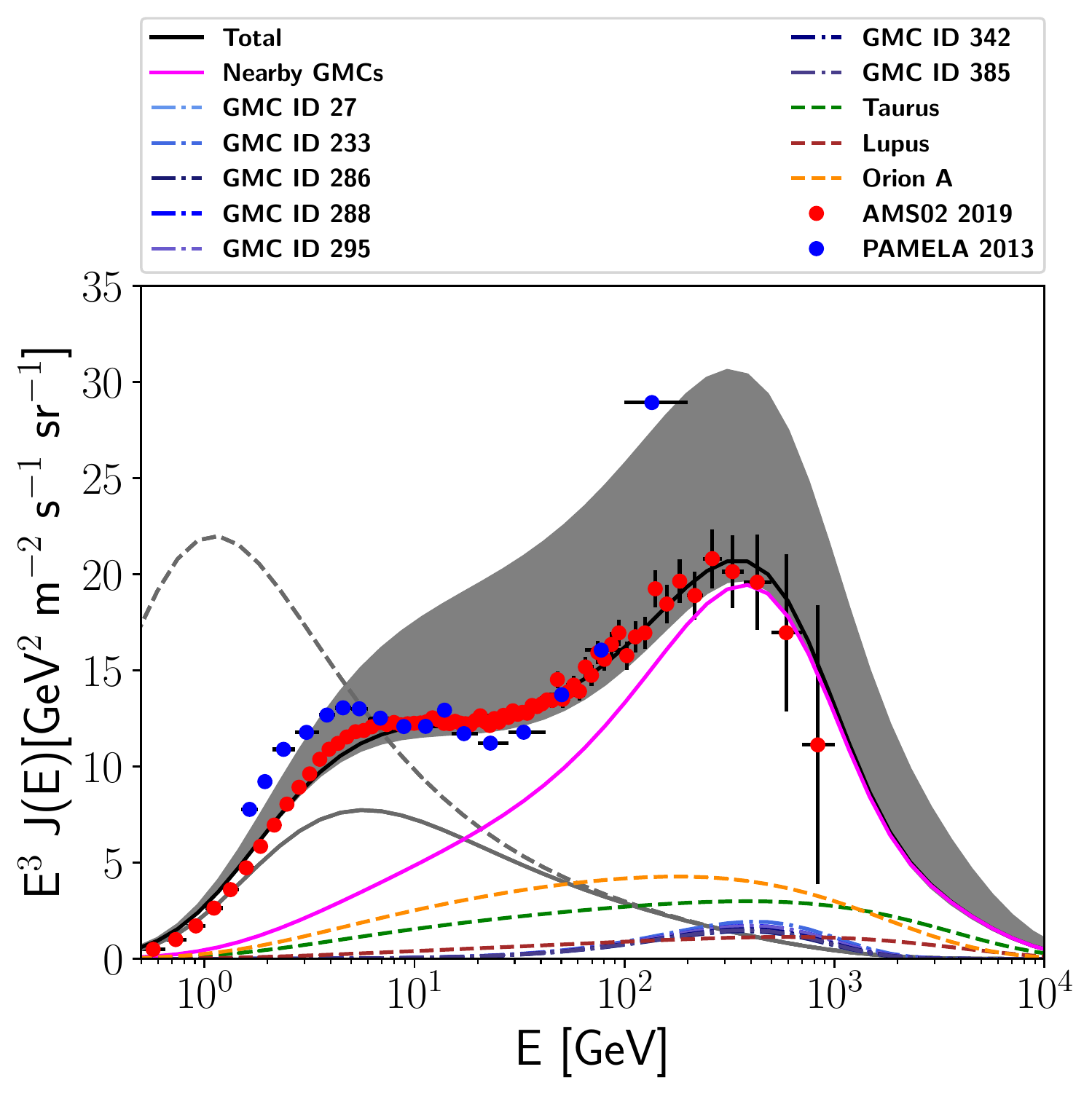}
\includegraphics[width=0.5\textwidth,origin=c,angle=0]{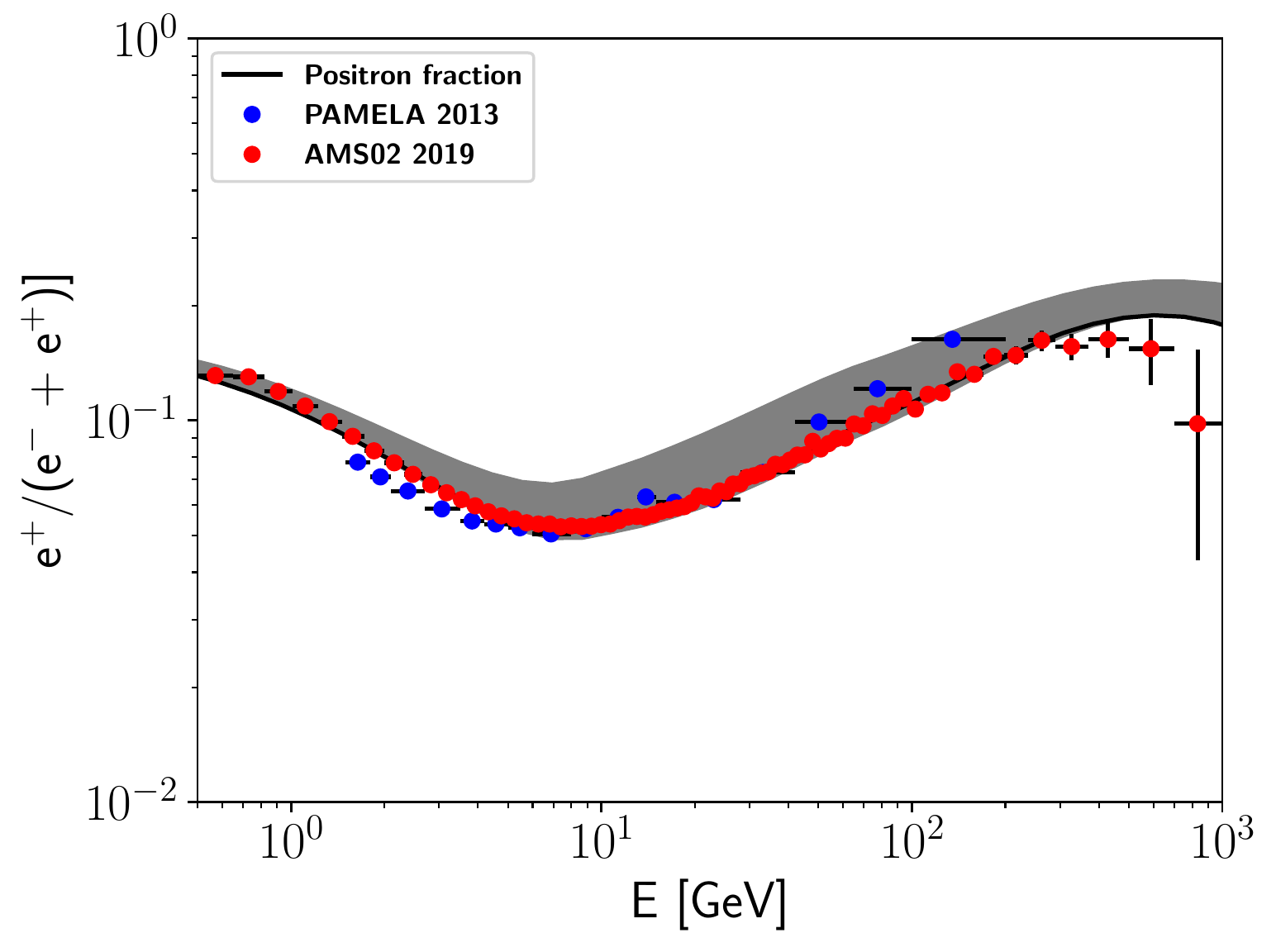}
\includegraphics[width=0.5\textwidth,origin=c,angle=0]{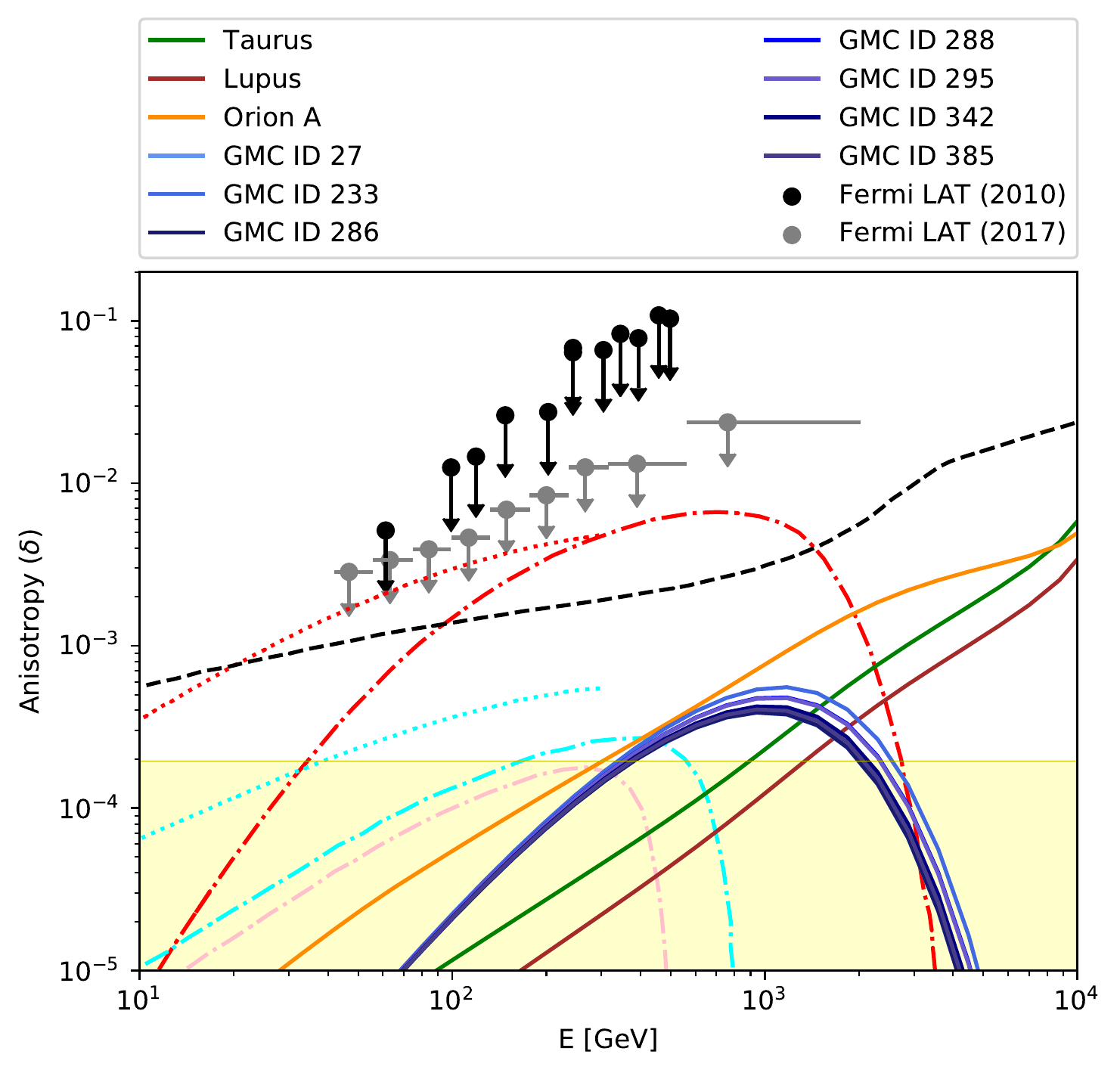}
\caption{\label{fig4} \textit{Upper Left :} Electron spectra fit from our model against AMS-02 \cite{aguilar19a} and PAMELA \cite{adriani11datab} data, \textit{Upper Right :} Positron spectra fit from our model against AMS-02 \cite{aguilar19b} and PAMELA \cite{adriani13dataa} data, \textit{Lower Left :} Positron fraction fit from our model against AMS-02 \cite{aguilar19b} and PAMELA \cite{adriani13dataa} data and \textit{Lower Right :} Anisotropies of GMCs and pulsars used to explain the positron excess, alongwith \textit{Fermi-}LAT anisotropy upper limits \cite{ackermann10,abdollahi17}. The gray shaded region signifies uncertainties due to variation of propagation parameters and uncertaintites in the normalization of nearby GMC fluxes.}
\end{figure} 

\section{Conclusion}
In this work, we have discussed an alternative, self-consistent scenario, where CR positrons are produced from CR interactions in the nearby GMCs and contribute significantly to the observed positron excess. Data of proton, antiproton spectra and B/C and $^{10}$Be\,/\,$^9$Be ratios as well as electron, positron spectra and positron fraction measured by AMS-02 and PAMELA, are fitted by our model quite well. Thus we conclude that nearby GMCs can play an important role in the observed positron excess. Further observations of the nearby GMCs by \textit{Fermi}-LAT, HESS, HAWC, LHAASO, CTA etc, can eluminate on the possibility of GMCs being an important contributors to the observed CR spectra.

%
%
%

\end{document}